# A Route to Nonrelativistic Altermagnetic Spin Splitting via Ultrafast Light

**Huang-Zhao-Xiang Chen[1,2], Lin-Ding Yuan[3], Wen-Hao Liu[1], Lin-Wang Wang[1], Jun-Wei Luo[1,2,*], and Zhi Wang[1,**]**

*[1] State Key Laboratory of Semiconductor Physics and Chip Technologies, Institute of Semiconductors, Chinese Academy of Sciences, Beijing, China.*

*[2] Center of Materials Science and Optoelectronics Engineering, University of Chinese Academy of Sciences, Beijing, China.*

*[3] Department of Materials Science and Engineering, Northwestern University, Evanston, Illinois 60208, USA*

**Contact authors:** * jwluo@semi.ac.cn; ** wangzhi@semi.ac.cn.

We identify a nonequilibrium route for generating altermagnetic spin splitting in antiferromagnet by ultrafast light. Unlike existing strategies, this route does not require relativistic angular-momentum transfer, static symmetry breaking, or auxiliary external fields. Using real-time time-dependent density functional theory, we demonstrate in the antiferromagnetic perovskite $KNiF_3$ that linearly polarized light can induce momentum-dependent altermagnetic spin splitting by breaking the effective time-reversal symmetry through photoexcited charge redistribution and the resulting lattice distortion. We provide a general symmetry selection rule for this route. These results establish a mechanism for ultrafast control of altermagnetism and extend its material realization into the nonequilibrium regime.

Magnetic order with zero net magnetization yet intrinsic momentum-dependent spin splitting even in the absence of relativistic effect (spin-orbit coupling, SOC) has recently drawn significant interest[1–5], known as altermagnetism (ALM)[6] or, more generally, nonrelativistic spin-splitting (NRSS)[7] magnetism. In these systems, the lifting of spin degeneracy originates from crystal and magnetic symmetries that break the effective time-reversal variances (combined space-time-reversal, $\mathcal{PT}$, and translational-spin-reversal, $\tau\mathcal{U}$))[7–9]. Altermagnets therefore provide a different realization of spin polarization distinct from both ferromagnets and relativistic spin-splitting materials.

Despite rapid theoretical progress, experimentally confirmed altermagnets remain limited to a small number of materials, including $RuO_2$[10–13], MnTe[14–16], $MnTe_2$[17], CrSb[18–20], and $CoNb_4Se_8$[21–24]. Recent efforts have explored several routes to generate or control altermagnetism. One route relies on static symmetry breaking, where auxiliary external electric fields or intrinsic polarizations lift Kramers degeneracy and enable nonrelativistic spin splitting[25–29], as developed later as altermagnetic multiferroicity[30–35]. Another route exploits relativistic mechanisms, in which SOC allows angular-momentum transfer from circularly polarized photon or chiral phonon to electronic spins, typically requiring heavy elements such as Pt, Ir, Tb, or Gd[36–39]. The requirements of multiferroicity or heavy elements substantially limit the range of candidate systems and manipulation strategy of altermagnetism.

This raises a fundamental question: can altermagnetic spin splitting be generated by a purely nonrelativistic and dynamical mechanism? In particular, it remains unknown whether light alone, without SOC or pre-existing symmetry breaking, can induce an altermagnetic state. Resolving this question would establish a distinct route to enable ultrafast control of magnetic states in solids.

Here we demonstrate such a mechanism. Using real-time time-dependent density functional theory (rt-TDDFT) combined with symmetry analysis, we show that linearly polarized light can drive an ultrafast transition from antiferromagnetism to altermagnetism in the absence of SOC. The mechanism arises directly from the *lattice* distortion induced by the photoexcited carrier occupation, which breaks the effective time-reversal symmetries. It leads to a nonequilibrium altermagnetic state with k-dependent spin splitting, persisting throughout the lifetime of the photoexcited carriers.

The mechanism is schematically illustrated in Fig. 1a. Electrons initially occupying the ground state are optically excited to higher states. This excitation transfers the system from a ground-state single-valley adiabatic potential energy surface (PES) to an excited-state PES featuring a multi-valley profile. On the ground-state PES, the minimum is antiferromagnetic with no spin splitting, whereas the minima on the excited-state PES correspond to structures with symmetry breaking required for altermagnetic order. Following photoexcitation, the system spontaneously evolves toward one of the symmetry-broken minima. Such photo-induced lattice distortion has been extensively reported in perovskites titanate, manganate, and ferrite[40–44]. We note that the symmetry conditions for altermagnetism, i.e., the breaking of $\mathcal{PT}$ and $\tau\mathcal{U}$ and the preservation of rotational-spin-reversal ($\mathcal{RU}$), are all induced by the lattice distortion and the resulting spatial symmetry breaking, without involving SOC or direct angular momentum transfer.

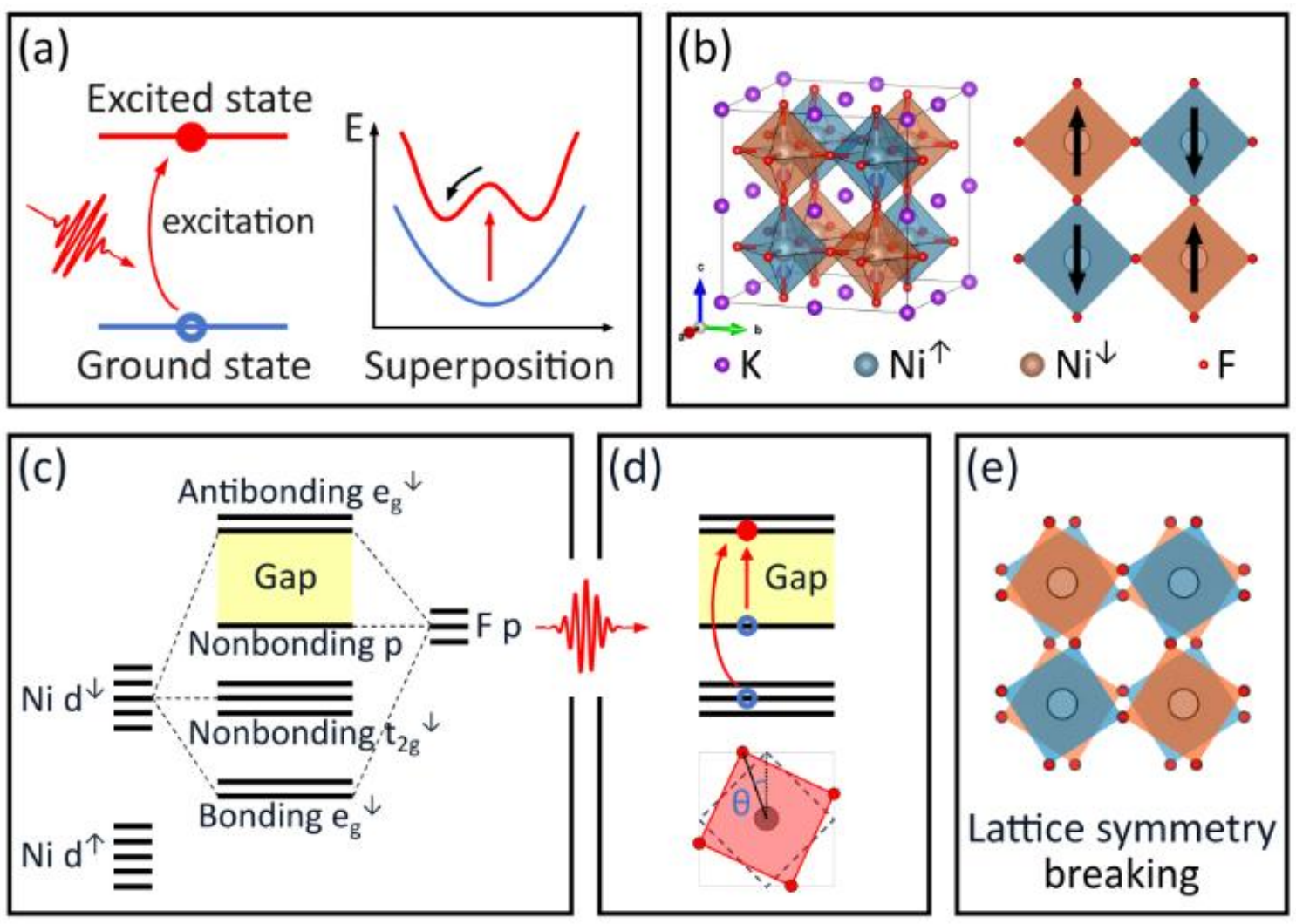

**FIG 1.** (a) Schematic illustration of light-induced symmetry breaking. (b) Crystal structure of ground-state $KNiF_3$, where opposite spin sublattices are represented by distinct octahedra. (c) Ligand-field splitting of Ni and F orbitals. (d) Photoexcitation process and the resulting sublattice rotation, characterized by the rotation angle θ. (e) Out-of-phase octahedral rotation (-), which breaks the lattice symmetry and enables altermagnetic order.

To demonstrate this mechanism, we first turn to a prototypical system, the G-type antiferromagnet $KNiF_3$[45,46]. In the ground state, $KNiF_3$ has a cubic lattice of a 2×2×2 supercell with collinear antiferromagnetic order, distinguished by orange and blue octahedra (Fig. 1b). It is a type-IV antiferromagnet[47] with no spin splitting. The ligand-field splitting from Ni-F bond lifts the degeneracy of Ni's 3d and F's p orbitals into bonding $e_g$, nonbonding $t_{2g}$, nonbonding p (from F), and antibonding $e_g$ orbitals (Fig. 1c). An ultrafast, near-band-gap optical excitation populates antibonding $e_g$ orbitals, weakening the Ni-F bonds and driving bond elongation. The photoexcited carriers can persist on picosecond timescales, whereas homogeneous lattice expansion under a typical laser spot would require nanosecond timescales. As a result, the lattice responds under an effectively constant-volume constraint. To accommodate the bond elongation within this constraint, the octahedra undergo rotations (Fig. 1d). Such rotations can break the symmetries that protect the Kramers degeneracy and result in an altermagnetic phase. For example, with the $a^0b^0c^-$ distortion mode where the rotations of octahedra in neighbor layers along [001] are antiparallel to each other (Fig. 1e), the effective time-reversal ($\mathcal{PT}$ and $\mathcal{U}\tau_{1/2}$) symmetries that link the spin-up and spin-down octahedra are broken, whereas the $m_{100}\mathcal{U}\tau_{1/2}$, $m_{010}\mathcal{U}\tau_{1/2}$, $m_{110}\mathcal{U}\tau_{1/2}$, and $m_{1\bar{1}0}\mathcal{U}\tau_{1/2}$ (combined with (100)-mirror, (010)-mirror, (110)-mirror, and $(1\bar{1}0)$-mirror reflections) are preserved. This will lead to a g-wave altermagnet[48]. Similarly, the $a^0b^-c^-$ distortion mode will lead to a d-wave altermagnet. Details are provided in Supplemental Materials Section II.

We now examine the photoexcitation dynamics using rt-TDDFT simulations. The results are presented in Fig. 2. $KNiF_3$ is driven by a linearly polarized laser pulse with central frequency 4.96 eV, comparable to the ground-state band gap (4.36 eV). We note that this gap is underestimated by DFT+U compared to experiment (~6 eV). However, since the mechanism is governed by the photoexcited carrier redistribution rather than the precise excitation energy, this underestimation does not qualitatively affect the results. The laser polarization is along the [111] direction. The Néel vector $\boldsymbol{L}$ is along [100]. Spin-orbit coupling is neglected in all simulations. Computational details are provided in Supplemental Material Section I.

The lattice structure exhibits pronounced octahedral rotation after the laser. At 410 fs (Fig. 2a), the rotation angles reach approximately 2.2°, 1.1°, and 12.3° about the three Cartesian axes. The resulting altermagnetic state is reflected in the spin splitting at the valence band maximum (VBM) in the $k_z$=0 plane (Fig. 2c-e). At early times (100 fs), the spin-splitting map shows a d-wave pattern, arising from a transient Jahn-Teller-like distortion. As the lattice dynamics becomes dominated by octahedral rotations, the system first develops the $a^0b^0c^-$ mode, leading to a g-wave spin-splitting pattern[48] at 410 fs. At later times (~900 fs), the rotation evolves into an $a^0b^-c^-$ mode, accompanied by the reappearance of d-wave features. Although both early- and late-time responses exhibit d-wave symmetry, they originate from distinct lattice distortions, as reflected by their difference with respect to the angular pattern (differ by π/4). To quantify this evolution, we project the spin-splitting map onto symmetry-adapted basis functions (Fig. 2b), revealing a clear switching between dominant d-wave and g-wave components, consistent with the k-space maps. Contributions from higher-order symmetries are negligible (see also Supplemental Materials Section III). The connection between lattice distortion modes and spin-splitting patterns is discussed below.

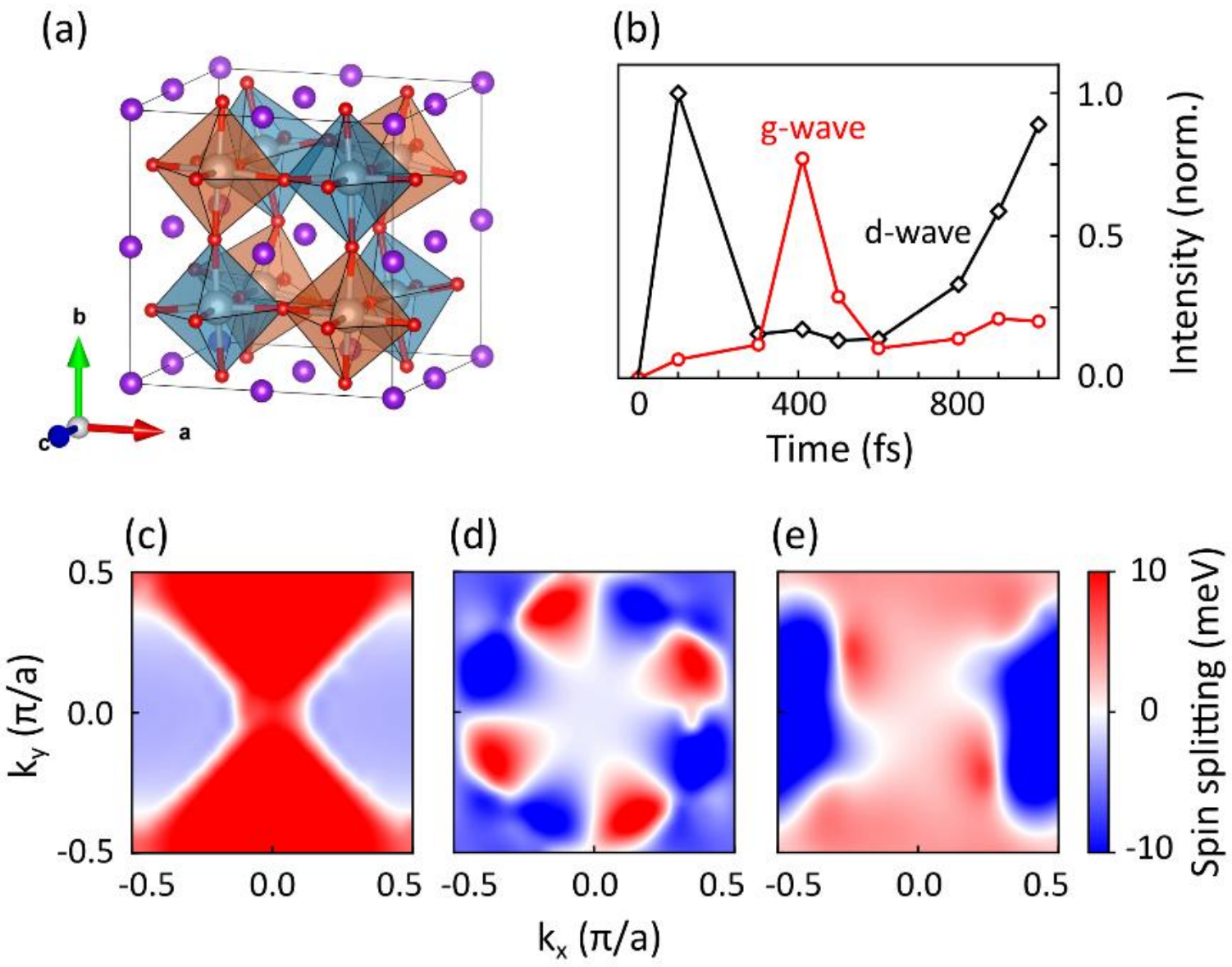

**FIG 2.** (a) $KNiF_3$ structure at 410 fs after photoexcitation, showing pronounced octahedral rotation. (b) Time evolution of the d-wave and g-wave components of the spin splitting at the valence band

maximum in the $k_z$=0 plane. (c-e) Spin-splitting maps at the valence band maximum in the $k_z$=0 plane at 100 fs, 410 fs, and 900 fs, respectively.

Beyond spin-splitting map, the emergent anomalous Hall conductivity (AHC) provides another direct, observable signature of altermagnetic state in this compound, which could be experimentally verified by time-resolved magneto-optical Kerr effect (tr-MOKE) and terahertz emission spectroscopy[49,50]. In Fig. 3, we show the time evolution of AHC coefficient. The ground state has zero AHC, while after the pump laser, $\sigma_{xy}$ shows prominent peaks near the VBM (Fig. 3a). The amplitudes of the largest $\sigma_{xy}$ in valence bands (labeled VB in Fig. 3b) oscillate between ±400 S/cm and persist throughout the simulation time (1 ps). We emphasize that, although the evaluation of AHC requires inclusion of SOC, the spin splitting and altermagnetism arise from lattice symmetry breaking without the need of SOC.

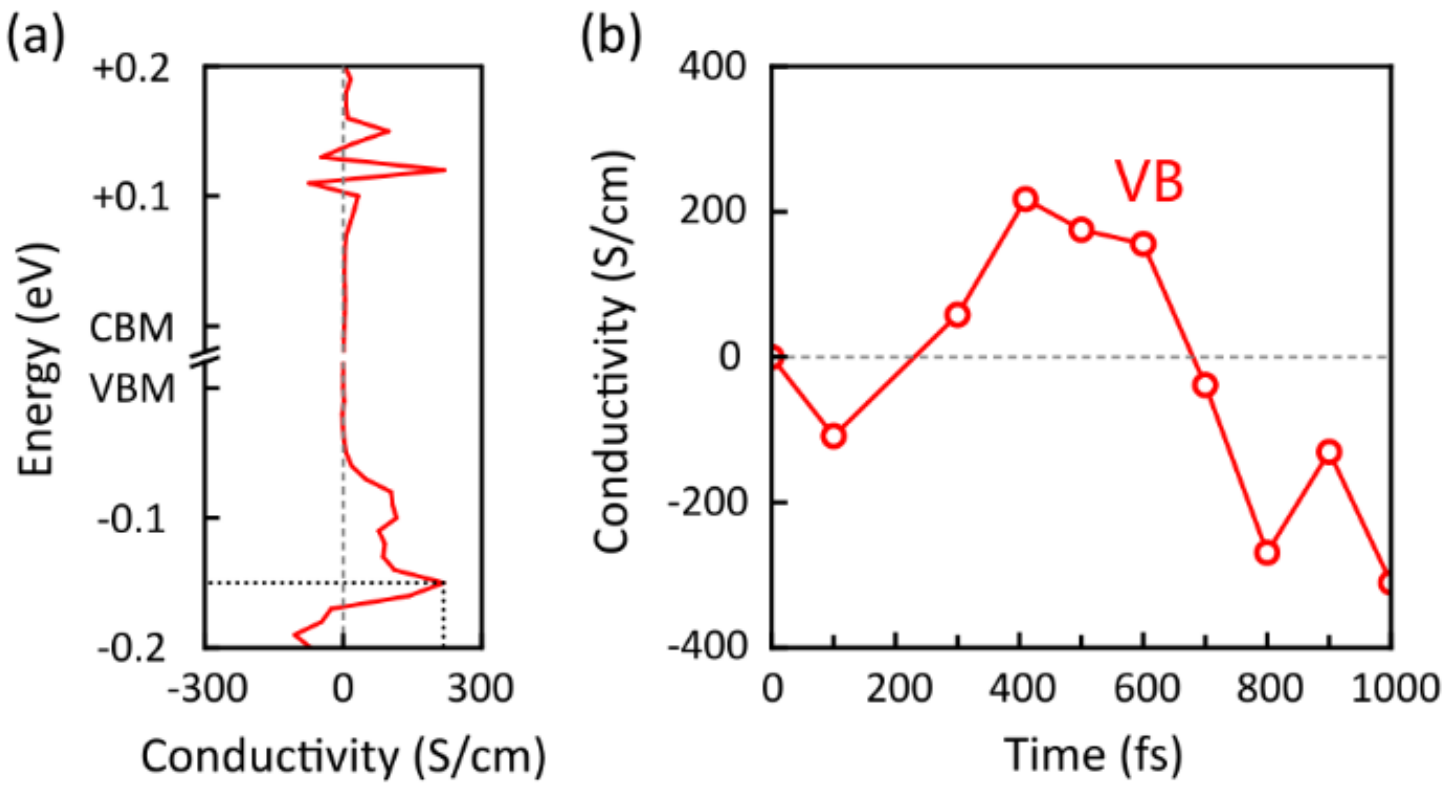

**FIG 3.** (a) Anomalous Hall conductivity $\sigma_{xy}$ at 410 fs after photoexcitation. Dash lines mark the peak in valence bands. (b) Time evolution of $\sigma_{xy}$ peak in valence bands.

To further uncover the microscopic origin of the lattice symmetry breaking, we analyze the electronic occupation and lattice response. The projected density of states (PDOS) shows that VBM and conduction band minimum (CBM) are dominated by Ni $t_{2g}$ and $e_g$ orbitals, respectively (Fig. 4a). Upon photoexcitation, electrons are promoted to antibonding $e_g$ states with an excitation population of 1.02% per formula unit (Fig. 4b). This nonequilibrium occupation drives a pronounced bond response. While equilibrium molecular dynamics (MD) shows only small fluctuations, rt-TDDFT reveals a substantial elongation of Ni-F bonds to ~1.94 Å at ~400 fs with oscillatory relaxation throughout the simulation time (Fig. 4c).

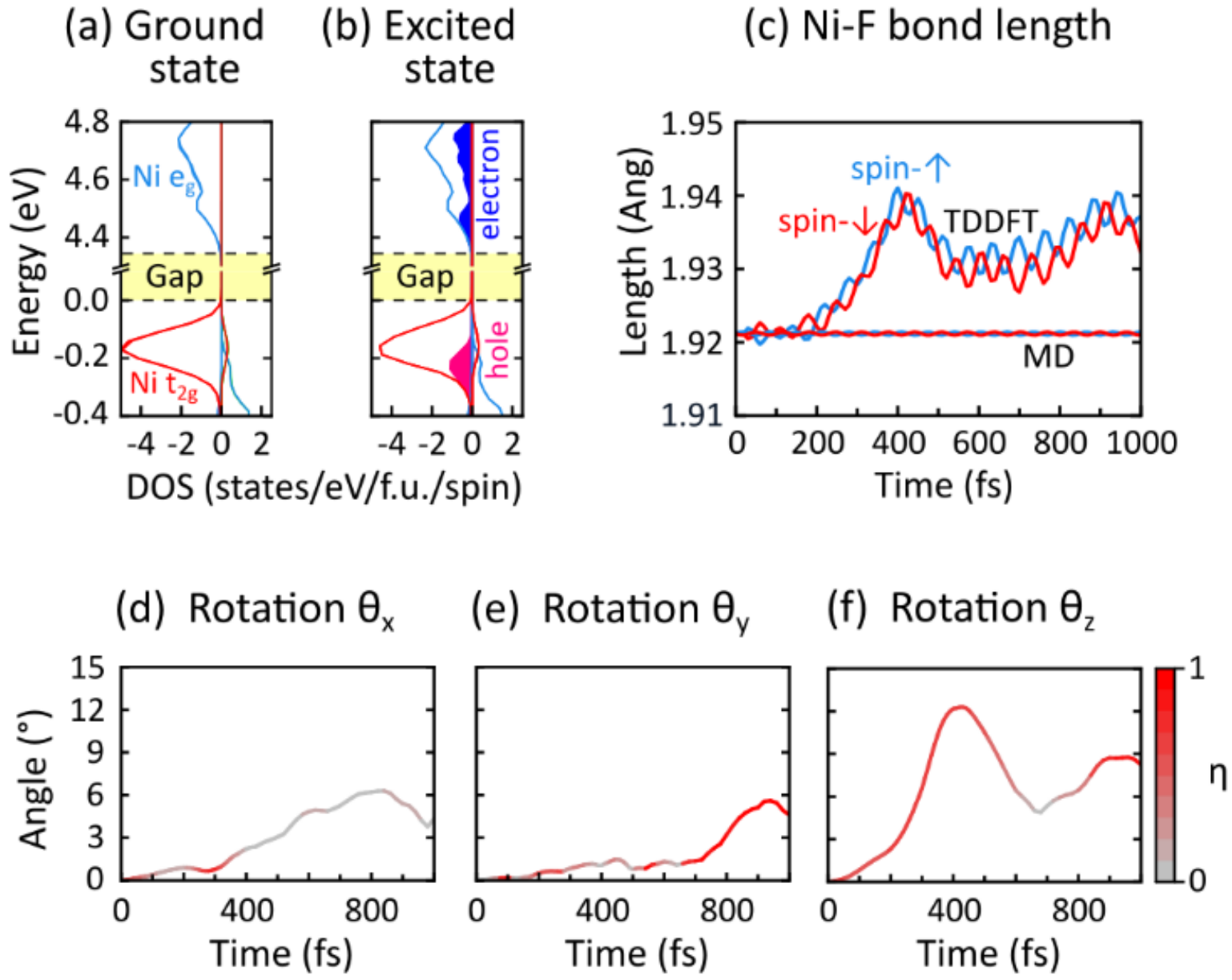

**FIG 4.** (a,b) Projected density of states in the ground and photoexcited states, showing carrier redistribution into antibonding states and hole formation near the valence band maximum. The positive and negative channels are spin-majority and spin-minority channels, respectively[51]. (c) Time evolution of the Ni-F bond length, comparing rt-TDDFT under photoexcitation with equilibrium molecular dynamics (MD). (d-f) Time evolution of the octahedral rotation angles θ about the $x$, $y$, and $z$ axes. The corresponding symmetry-breaking parameter $\eta$ quantifies the degree of effective time-reversal symmetry breaking.

To quantify the rotation-induced symmetry breaking, we introduce a dimensionless parameter,

$$\eta_\alpha = 1 - \frac{|\theta_\alpha^u - \theta_\alpha^l|}{2\,\max\left(|\theta_\alpha^u|, |\theta_\alpha^l|\right)}$$

where $\alpha = x, y, z$, and $\theta_\alpha^{u(l)}$ denotes the octahedral rotation angle of the upper (lower) layer along the α direction. By construction, $\eta = 0$ corresponds to pure in-phase rotation preserving effective time-reversal symmetries, while $\eta = 1$ is the pure out-of-phase rotation associated with the strongest symmetry breaking. The time evolutions of the averaged rotation angle $\theta_\alpha(t) = \text{average}\left(\left|\theta_\alpha^{u(l)}(t)\right|\right)$ as well as the $\eta_\alpha$ are shown in Fig. 4d-f. Symmetry breaking from rotations about $x$ axis remains negligible throughout the dynamics. Rotations about $z$ axis dominate in the early stage (0-600 fs), where $\eta_z \approx 1$ and the maximum rotation angle is ~12° at ~400 fs. This corresponds to the $a^0b^0c^-$ distortion. We have also confirmed that such out-of-phase mode is more energetically favored on the excited PES than the in-phase mode (see also Supplemental Materials Section IV). The simultaneous emergence of maximal $\theta_z$, dominant g-wave spin splitting (Fig. 2b), and peak AHC $\sigma_{xy}$ (Fig. 3b) establishes a direct link between out-of-phase rotation and the altermagnetic response. Around 700 fs, $\eta_z$ decreases substantially, accompanied by the suppression of the AHC. At later times, rotations about $y$ axis become significant with $\eta_y \approx 1$, giving rise to the $a^0b^-c^-$ mode and an enhanced $\sigma_{xy}$. The switching between these distortion modes accounts for the evolution of the spin-splitting patterns observed in Fig. 2b-e. This establishes a direct structure-symmetry-response connection underlying the light-induced altermagnetic state.

We now formulate the selection rules governing light-induced altermagnetism. The driving force on a distortion mode $\nu$ is given by $F_\nu \sim \mathrm{Tr}\left[\hat{g}_\nu \delta\hat{\rho}^{(2)}\right]$, where $\hat{g}_\nu$ is the electron-phonon coupling operator and $\delta\hat{\rho}^{(2)}$ is the bilinear density response to the laser field. This leads to two selection rules. **(1) Symmetry selection rule.** The driven phonon mode must belong to the symmetric bilinear channel of the laser field, $\Gamma_\nu \subset \mathrm{Sym}^2(\Gamma_{laser})$. **(2) Momentum selection rule.** Optical excitation predominantly involves vertical transitions in $k$-space, requiring the driven phonon mode to have vanishing wave vector, $\boldsymbol{q} = 0$. In $KNiF_3$, the AFM structure (i) reduces the unitary little co-group from $O_h$ to $D_{4h}$, and (ii) folds zone-boundary phonon modes to the zone center. With Néel vector $\boldsymbol{L} \parallel [100]$ and G-type AFM order, the out-of-phase rotation becomes a $\boldsymbol{q} = 0$ mode and decomposes as $A_{2g} \oplus E_g$, corresponding to the longitudinal ($R_\parallel \sim A_{2g}$, parallel to $\boldsymbol{L}$) and transverse ($R_\perp \sim E_g$, perpendicular to $\boldsymbol{L}$) components. The electric field transforms as $\Gamma_{laser} = A_{2u} \oplus E_u$, yielding $\mathrm{Sym}^2(A_{2u} \oplus E_u) = A_{1g} \oplus B_{2g} \oplus E_g$. Thus, only the transverse $E_g$ components are symmetry-allowed, while the longitudinal $A_{2g}$ component is forbidden. This explains the selective activation of the $a^0b^0c^-$ rotation at early times (Fig. 4d-f), and implies that light-induced altermagnetism emerges only when the laser polarization is not parallel to the Néel vector. This principle applies generally to cubic G-type AFM perovskites, including $SrMnO_3$, $PbCrO_3$, $RbMnF_3$, and $KCoF_3$. Details are provided in Supplemental Material Section V.

The mechanism identified here is distinct from relativistic angular-momentum transfer or static symmetry breaking. Instead, it originates from symmetry-selective lattice distortions driven by photoexcited carrier populations. Our selection rule establishes a general symmetry criterion for realizing such light-induced altermagnetism. Our findings provide a route toward ultrafast control of altermagnetism and extend its material realization into the nonequilibrium regime.

## Reference

1. Hayami, S., Yanagi, Y. & Kusunose, H. Momentum-Dependent Spin Splitting by Collinear Antiferromagnetic Ordering. *J. Phys. Soc. Jpn.* **88**, 123702 (2019).
2. Šmejkal, L., Sinova, J. & Jungwirth, T. Emerging Research Landscape of Altermagnetism. *Phys. Rev. X* **12**, 040501 (2022).
3. Yuan, L.-D., Wang, Z., Luo, J.-W., Rashba, E. I. & Zunger, A. Giant momentum-dependent spin splitting in centrosymmetric low- Z antiferromagnets. *Phys. Rev. B* **102**, 014422 (2020).
4. Šmejkal, L., González-Hernández, R., Jungwirth, T. & Sinova, J. Crystal time-reversal symmetry breaking and spontaneous Hall effect in collinear antiferromagnets. *Sci. Adv.* **6**, eaaz8809 (2020).
5. Mazin, I. I., Koepernik, K., Johannes, M. D., González-Hernández, R. & Šmejkal, L. Prediction of unconventional magnetism in doped $FeSb_2$. *Proc. Natl. Acad. Sci. U.S.A.* **118**, e2108924118 (2021).
6. Yuan, L.-D., Georgescu, A. B. & Rondinelli, J. M. Nonrelativistic Spin Splitting at the Brillouin Zone Center in Compensated Magnets. *Phys. Rev. Lett.* **133**, 216701 (2024).
7. Yuan, L.-D., Wang, Z., Luo, J.-W. & Zunger, A. Prediction of low-Z collinear and noncollinear antiferromagnetic compounds having momentum-dependent spin splitting even without spin-orbit coupling. *Phys. Rev. Materials* **5**, 014409 (2021).
8. Liu, P., Li, J., Han, J., Wan, X. & Liu, Q. Spin-Group Symmetry in Magnetic Materials with Negligible Spin-Orbit Coupling. *Phys. Rev. X* **12**, 021016 (2022).
9. Zhang, X., Xiong, J.-X., Yuan, L.-D. & Zunger, A. Prototypes of Nonrelativistic Spin Splitting and Polarization in Symmetry Broken Antiferromagnets. *Phys. Rev. X* **15**, 031076 (2025).
10. Liu, J. *et al.* Absence of Altermagnetic Spin Splitting Character in Rutile Oxide RuO 2. *Phys. Rev. Lett.* **133**, 176401 (2024).
11. Šmejkal, L. *et al.* Chiral Magnons in Altermagnetic RuO 2. *Phys. Rev. Lett.* **131**, 256703 (2023).
12. Fedchenko, O. *et al.* Observation of time-reversal symmetry breaking in the band structure of altermagnetic $RuO_2$. *Sci. Adv.* **10**, eadj4883 (2024).

13. Ho, D. Q., To, D. Q., Hu, R., Bryant, G. W. & Janotti, A. Symmetry-breaking induced surface magnetization in non-magnetic RuO$_2$. *Phys. Rev. Materials* **9**, 094406 (2025).

14. Devaraj, N., Bose, A. & Narayan, A. Interplay of altermagnetism and pressure in hexagonal and orthorhombic MnTe. *Phys. Rev. Materials* **8**, 104407 (2024).

15. Liu, Z., Ozeki, M., Asai, S., Itoh, S. & Masuda, T. Chiral Split Magnon in Altermagnetic MnTe. *Phys. Rev. Lett.* **133**, 156702 (2024).

16. Krempaský, J. *et al.* Altermagnetic lifting of Kramers spin degeneracy. *Nature* **626**, 517–522 (2024).

17. Zhu, Y.-P. *et al.* Observation of plaid-like spin splitting in a noncoplanar antiferromagnet. *Nature* **626**, 523–528 (2024).

18. Ding, J. *et al.* Large Band Splitting in g -Wave Altermagnet CrSb. *Phys. Rev. Lett.* **133**, 206401 (2024).

19. Reimers, S. *et al.* Direct observation of altermagnetic band splitting in CrSb thin films. *Nat Commun* **15**, 2116 (2024).

20. Santhosh, S. *et al.* Altermagnetic Band Splitting in 10 nm Epitaxial CrSb Thin Films. *Advanced Materials* **37**, e08977 (2025).

21. Mandujano, H. C. *et al.* Evolution of Altermagnetism to Spin Density Waves in Co$_x$ NbSe$_2$. *J. Am. Chem. Soc.* **147**, 44926–44940 (2025).

22. Regmi, R. B. *et al.* Altermagnetism in the layered intercalated transition metal dichalcogenide CoNb4Se8. *Nat Commun* **16**, 4399 (2025).

23. Sah, A. *et al.* Altermagnetism, Kagome Flat Band, and Weyl Fermion States in Magnetically Intercalated Transition Metal Dichalcogenides. Preprint at https://doi.org/10.48550/arXiv.2510.21968 (2025).

24. Graham, J. N. *et al.* Local probe evidence supporting altermagnetism in Co$_{1/4}$NbSe$_2$. Preprint at https://doi.org/10.48550/arXiv.2503.09193 (2025).

25. Bai, L. *et al.* Altermagnetism: Exploring New Frontiers in Magnetism and Spintronics. *Adv Funct Materials* **34**, 2409327 (2024).

26. Mazin, I., González-Hernández, R. & Šmejkal, L. Induced Monolayer Altermagnetism in MnP(S,Se)$_3$ and FeSe. Preprint at https://doi.org/10.48550/arXiv.2309.02355 (2023).
27. Zhou, P. *et al.* Transition from antiferromagnets to altermagnets: Symmetry-Breaking Theory. *Phys. Rev. B* **112**, 144419 (2025).
28. Spaldin, N. A. Multiferroics beyond electric-field control of magnetism. *Proc. R. Soc. A.* **476**, 20190542 (2020).
29. Spaldin, N. A. & Ramesh, R. Advances in magnetoelectric multiferroics. *Nature Mater* **18**, 203–212 (2019).
30. Chen, Y., Liu, X., Lu, H.-Z. & Xie, X. C. Electrical Switching of Altermagnetism. *Phys. Rev. Lett.* **135**, 016701 (2025).
31. Sun, W., Yang, C., Wang, X., Huang, S. & Cheng, Z. Altermagnetic multiferroics with symmetry-locked magnetoelectric coupling. *Nat. Mater.* https://doi.org/10.1038/s41563-026-02518-5 (2026) doi:10.1038/s41563-026-02518-5.
32. Gu, M. *et al.* Ferroelectric Switchable Altermagnetism. *Phys. Rev. Lett.* **134**, 106802 (2025).
33. Duan, X. *et al.* Antiferroelectric Altermagnets: Antiferroelectricity Alters Magnets. *Phys. Rev. Lett.* **134**, 106801 (2025).
34. Ding, N., Ye, H., Wang, S.-S. & Dong, S. Ferroelastically tunable altermagnets. *Phys. Rev. B* **112**, L220410 (2025).
35. Wan, X., Mandal, S., Guo, Y. & Haule, K. High-Throughput Search for Metallic Altermagnets by Embedded Dynamical Mean Field Theory. *Phys. Rev. Lett.* **135**, 106501 (2025).
36. Li, G. *et al.* Ultrafast kinetics of the antiferromagnetic-ferromagnetic phase transition in FeRh. *Nat Commun* **13**, 2998 (2022).
37. Leenders, R. A., Afanasiev, D., Kimel, A. V. & Mikhaylovskiy, R. V. Canted spin order as a platform for ultrafast conversion of magnons. *Nature* **630**, 335–339 (2024).
38. Igarashi, J. *et al.* Optically induced ultrafast magnetization switching in ferromagnetic spin valves. *Nat. Mater.* **22**, 725–730 (2023).
39. Davies, C. S. *et al.* Phononic switching of magnetization by the ultrafast Barnett effect. *Nature*

**628**, 540–544 (2024).

40. Subedi, A., Cavalleri, A. & Georges, A. Theory of nonlinear phononics for coherent light control of solids. *Phys. Rev. B* **89**, 220301 (2014).
41. Först, M. *et al.* Nonlinear phononics as an ultrafast route to lattice control. *Nature Phys* **7**, 854–856 (2011).
42. Disa, A. S., Nova, T. F. & Cavalleri, A. Engineering crystal structures with light. *Nat. Phys.* **17**, 1087–1092 (2021).
43. Juraschek, D. M., Fechner, M. & Spaldin, N. A. Ultrafast Structure Switching through Nonlinear Phononics. *Phys. Rev. Lett.* **118**, 054101 (2017).
44. Song, C. *et al.* Electronic Origin of Laser-Induced Ferroelectricity in $SrTiO_3$. *J. Phys. Chem. Lett.* **14**, 576–583 (2023).
45. Ceriotti, D. *et al.* Mechanochemical synthesis of fluorinated perovskites $KCuF_3$ and $KNiF_3$. *RSC Mechanochem.* **1**, 520–530 (2024).
46. Markovin, P. A., Petrov, S. V. & Pisarev, R. V. Isotropic and anisotropic magnetic refraction of light in the antiferromagnets KNiF, and RbMnF,.
47. Bradley, C. J. & Cracknell, A. P. The mathematical theory of symmetry in solids;: Representation theory for point groups and space groups,. in (1972).
48. Ghorashi, S. A. A. & Li, Q. Dynamical Generation of Higher-order Spin-Orbit Couplings, Topology and Persistent Spin Texture in Light-Irradiated Altermagnets. *Phys. Rev. Lett.* **135**, 236702 (2025).
49. Schilberth, F. *et al.* Magneto-optical detection of topological contributions to the anomalous Hall effect in a kagome ferromagnet. *Phys. Rev. B* **106**, 144404 (2022).
50. Matsuda, T. *et al.* Ultrafast Dynamics of Intrinsic Anomalous Hall Effect in the Topological Antiferromagnet Mn 3 Sn. *Phys. Rev. Lett.* **130**, 126302 (2023).
51. Trimarchi, G., Wang, Z. & Zunger, A. Polymorphous band structure model of gapping in the antiferromagnetic and paramagnetic phases of the Mott insulators MnO, FeO, CoO, and NiO. *Phys. Rev. B* **97**, 035107 (2018).

## ACKNOWLEDGMENTS

This work is supported by the National Natural Science Foundation of China (12174380). Z.W. is supported by the CAS project for Young Scientists in Basic Research under Grant No. YSBR-030.

## DATA AVAILABILITY

All data presented in this study are available from the corresponding authors upon reasonable request.